# Probing electron spin dynamics in single telecom InAs(P)/InP quantum dots using the Hanle effect

Maja Wasiluk[1,a)], Helena Janowska[1], Anna Musiał[1], Johann P. Reithmaier[2], Mohamed Benyoucef[2], Wojciech Rudno-Rudziński[1]

[1]*Department of Experimental Physics, Faculty of Fundamental Problems of Technology, Wrocław University of Science and Technology, Wyb. Wyspiańskiego 27, 50-370 Wrocław, Poland*

[2]*Institute of Nanostructure Technologies and Analytics, CINSaT, University of Kassel, Heinrich-Plett-Str. 40, 34132 Kassel, Germany*

**ABSTRACT**

Spins of carriers confined in quantum dots (QDs) are promising candidates for qubits due to their relatively long spin relaxation times. However, the electron spin dephasing, primarily driven by hyperfine interactions with nuclear spins, can limit their coherence. Here, we report the first Hanle effect demonstration in single InAs(P)/InP QDs emitting in the telecom C-band leading to experimental determination of electron spin dephasing time. Using polarization-resolved photoluminescence spectroscopy, we identified excitonic complexes and confirmed the presence of a negatively charged trion, exhibiting a degree of circular polarization (DOCP) of −36% under quasi-resonant excitation. From the analysis of Hanle linewidth and employing a previously reported value of the electron g-factor, we extracted an electron spin dephasing time of $\mathrm{T}_2^* = 1.59 \pm 0.49\ \mathrm{ns}$. Despite the large indium nuclear spin, the obtained $\mathrm{T}_2^*$ is comparable to values reported for GaAs-based QDs, which we attribute to the larger volume of the InAs(P)/InP QDs. These findings confirm the potential of InP-based telecom QDs for use in spin–photon interfaces.

[a)]Author to whom correspondence should be addressed: maja.wasiluk@pwr.edu.pl

The requirement for long qubit lifetime and coherence time is the main driving force behind the search for suitable platforms for quantum information processing. Identifying mechanisms that limit the spin coherence can help improve qubit design and enhance their potential. Together with developed methods of extending and controlling their coherence, qubits can be implemented in quantum technology.[1] Over the years, there have been many proposals of platforms serving as qubits, but each comes with its limitations. Superconducting circuits offer mature, scalable architectures, but require operation at millikelvin temperatures, and are incompatible with photonic systems due to their microwave-frequency emission.[2] While trapped ions and Rydberg atoms exhibit very long dephasing times (exceeding 1 s),[3-5] their implementation is hindered by challenges in miniaturization and on-chip integration. Nuclear spin qubits benefit from exceptional dephasing times; however, they suffer from slow gate operations and difficult addressability.[6] Photonic qubits, on the other hand, are ideal for long-distance quantum communication, but lack efficient two-qubit gates due to weak photon-photon interactions, which makes them difficult to scale in logic-based architectures.[7-8] In contrast to other platforms, quantum dots (QDs) can be efficiently integrated with on-chip photonic structures and are compatible with standard semiconductor fabrication technologies. They also do not require complex experimental infrastructure. Moreover, QDs can operate in the telecom range, enabling compatibility with flying qubits for quantum communication. Carrier spins in QDs can also be addressed and controlled using purely optical methods. However, the interaction with nuclei is a strong decoherence mechanism for localized electrons.[9] A short spin coherence time limits the number of quantum operations that can be performed before the decoherence randomizes the spin direction. This leads to an increased error rate, causing the need for advanced error-correction protocols that require even more resources. Moreover, dephasing due to the hyperfine interaction has also been shown to reduce the fidelity of quantum light state generation,[10] thereby limiting their potential in spin-photon interfaces that support hybrid quantum architectures. Understanding and mitigating the underlying spin dephasing mechanisms, particularly those arising from nuclear spin fluctuations, is essential for developing robust carrier spin-based qubits.

Charged QDs are a preferable choice for the spin qubits because the residual charge spin relaxation time does not depend on the exciton lifetime. Through optical techniques, carrier spin states in semiconductor QDs can be initialized, coherently manipulated, and read out.[11-12] In addition, optical methods provide a platform for characterizing coherent properties of confined spins. Specifically, time-resolved optical techniques, such as Ramsey interference, Hahn echo, and pump-probe photoluminescence, allow for the direct measurement of carrier spin dephasing times.[11,13-14]

The abovementioned methods demand time-resolved optical setups and ultrashort (picosecond or femtosecond) laser pulses, with precise spectral and temporal shaping. This requires advanced equipment and complicates the whole experiment. An alternative is the Hanle effect measurement, in which an applied external magnetic field causes the depolarization of the trion emission in the low magnetic field regime (typically for values < 1 T, which can be achieved with electromagnetic coils). The electron spin dephasing time can be probed through its spin precession rate in an external magnetic field. The Hanle effect has been well studied for QD ensembles in many material systems with emission wavelengths below 1 µm, such as InAs/GaAs,[15] (In,Ga)As/GaAs,[16-17] InAlAs/AlGaAs,[18] and InP/GaAs,[19] as well as on the level of a single GaAs QD (a natural QD at the interface of a 4 nm GaAs/AlGaAs quantum well).[20] However, for QDs emitting in the telecom spectral range, the Hanle studies have only been performed on the InAs/InP QD ensemble.[21] There are only a few reports on the single spin dephasing study of carriers confined in QDs emitting in the telecom spectral range. A hole spin in a single InAs/GaAs QD was measured through the Ramsey interference.[11] Another recent report shows the study on the single carriers spins confined in the InGaAs/GaAs QDs grown on an InGaAs metamorphic buffer layer performed by measuring negative trion's Larmor precession under an externally applied in-plane magnetic field.[22] Additionally, spin dephasing in single InAs/InP QDs emitting in the C-band was recently investigated using polarization-resolved resonance fluorescence measurements.[23] We use here the Hanle effect measurement to study the single-spin dephasing time of the optically oriented residual electron in a QD. This is the first report demonstrating the Hanle effect on the single spin in QDs emitting in the telecom C-band.

In this study, we explore symmetric self-assembled InAs(P) QDs grown on an InP Fe-doped substrate via molecular beam epitaxy (MBE), assisted by the ripening process that led to QDs' low spatial density of ~$2 \times 10^9$ $cm^{-2}$.[24] QDs were embedded directly in the InP matrix and grown on top of a distributed Bragg reflector (25 pairs of InP/InAlGaAs), resulting in photon extraction efficiency of 13.3% into the numerical aperture (NA) of 0.4 for QDs in non-deterministically fabricated cylindrical mesas.[25] These QDs emit in the 3rd telecom window and exhibit a low probability of multiphoton events ($g^{(2)}(0) < 0.01$),[26] making them promising for generating non-classical light and, thus, optically driven spintronics compatible with the telecom infrastructure.

Here, we show the photoluminescence study of the single QD leading to the trion identification and the microphotoluminescence excitation (µPLE) measurement for evaluation of the higher energy states within the QD. The identification of excitonic complexes was performed using the excitation power-dependent and polarization-resolved photoluminescence under non-resonant excitation with a continuous-wave (CW) laser emitting at 640 nm. For quasi-resonant excitation, a CW external cavity semiconductor infrared laser tunable in the range of 1410–1535 nm was used in the µPLE measurements, as well as in the measurements of the degree of circular polarization and the Hanle curve. Polarization of both the excitation laser and the detected signal was controlled using a set of polarization optics. For Hanle measurements, the sample was mounted on an elongated cold finger of the cryostat, positioned between two electromagnet coils providing a magnetic field up to 400 mT in the Voigt configuration. This configuration was calibrated using an ammeter and a teslameter. The typical integration time per magnetic field point was 20 s, optimized to balance the signal-to-noise ratio and measurement time. In all the experiments, the emission from the QDs was collected with a microscope objective with an NA of 0.4 and a working distance of 20 mm. Emission was dispersed by a 1 m focal-length monochromator and detected using a liquid-nitrogen-cooled InGaAs multichannel linear array. The sample was kept at a temperature between 5 and 10 K, except during the Hanle effect measurements. Due to the greater distance between the cold finger and the sample during these measurements, the temperature was estimated to be around 15 K, based on the emission from the reference sample acting as a thermometer. Similar measurements were repeated on several QDs of the same type, yielding consistent results.

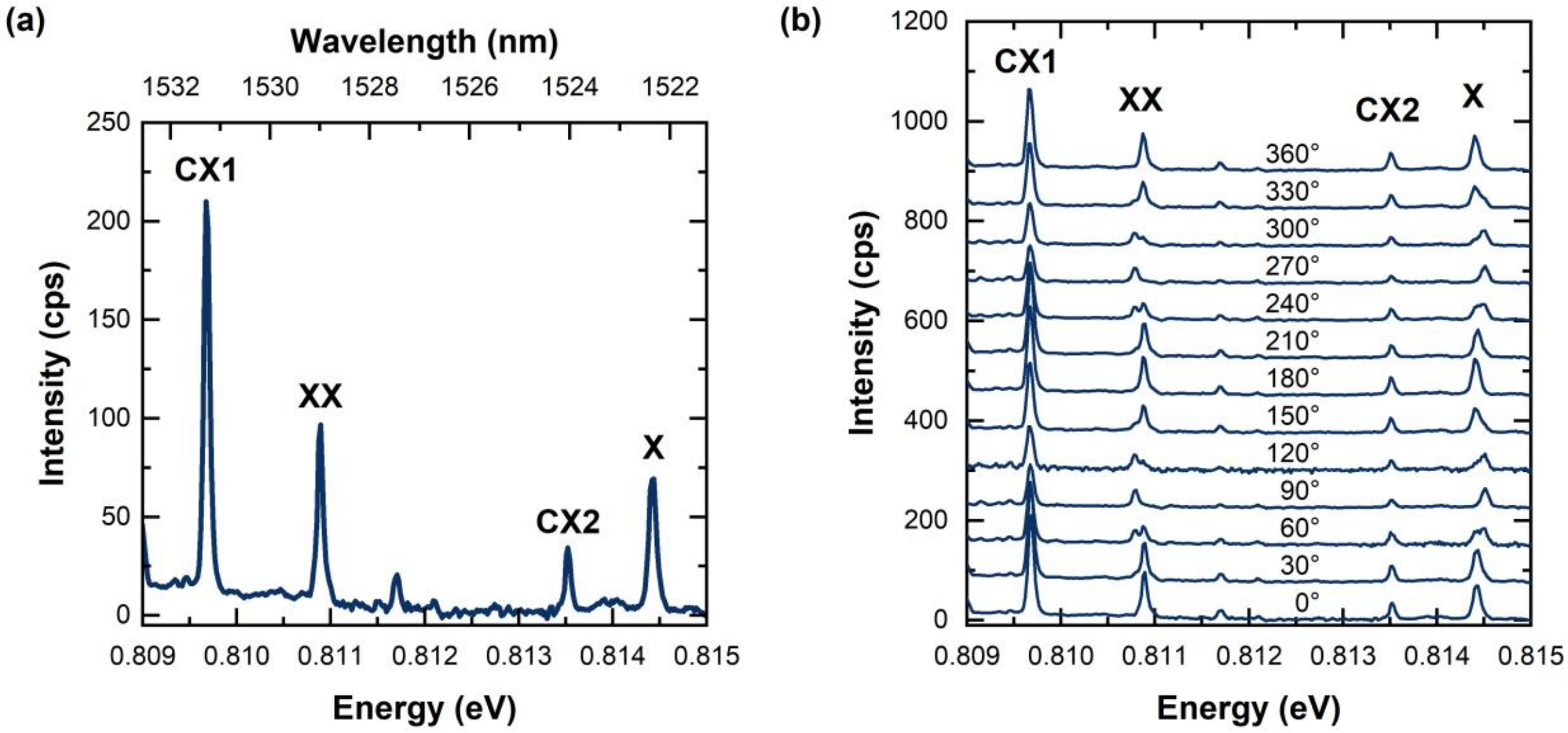

***FIG 1.*** *(a) Photoluminescence spectrum of studied QD under non-resonant excitation with excitation power of 25 μW. (b) Polarization-resolved spectra from the same QD under the same excitation conditions. Energies of X and XX lines change sinusoidally, implying splitting of 2 orthogonal states of neutral excitonic complex with two components visible at intermediate polarization angles. For CX1 and CX2, no splitting is observed.*

Power-dependent photoluminescence measurements were used as a preliminary method to pinpoint charged excitonic complexes (trions). Polarization-resolved photoluminescence (Fig. 1(b)) further supported the identification of CX1 and CX2 as trions. In the neutral excitonic states, fine structure splitting leads to orthogonally polarized doublets. Trion states lack such splitting due to the absence of the electron-hole exchange interaction of the single carrier. We attribute the CX1 and CX2 lines to trion recombination, and the X and XX lines to exciton and biexciton recombination, respectively. We expect trions to be mostly negatively charged due to the Fe doping in the substrate, making the structure n-type.

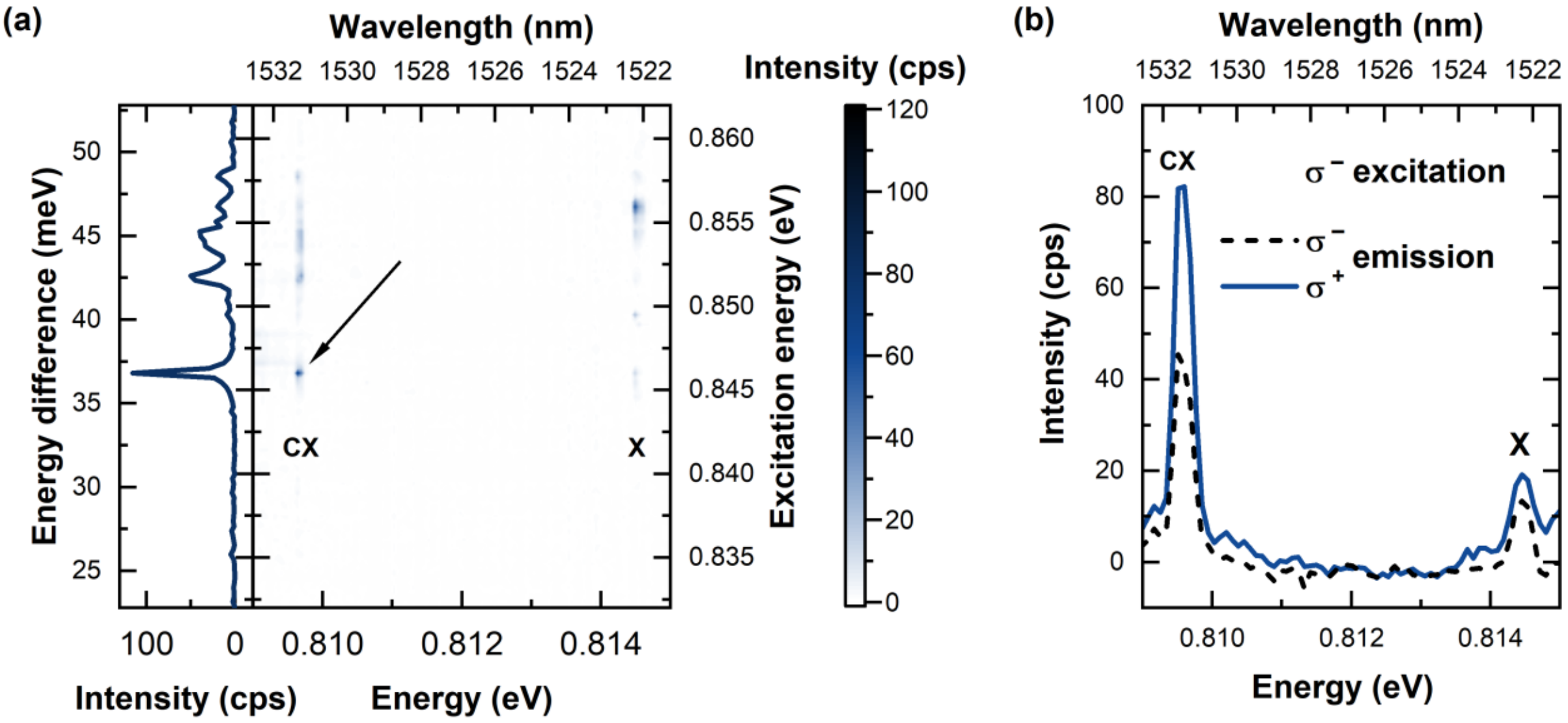

***FIG. 2.*** *(a) Photoluminescence excitation map. Arrow points at the resonance chosen for further studies. (b) PL spectra of CX1 and X line under quasi-resonant circularly-polarized excitation and circular-polarization-resolved detection. In contrast to X, CX1 line exhibits a degree of circular polarization of -36%.*

Due to the presence of residual carriers, whose spin dephasing time is not limited by the exciton recombination time, charged QDs are chosen as potential spin qubits. Studying and controlling the spin of carriers using optical methods requires certain excitation conditions. Non-resonant excitation, i.e. into the barrier/wetting layer, leads to the randomization of the carrier spin direction via multiple relaxation processes, including non-preserving spin scattering. In contrast, quasi-resonant excitation into the higher energy states within a QD assures a single relaxation process that typically preserves the spin orientation.[27] To determine the excitation energy closest to the emission energy, we performed the µPLE measurement. Fig. 2(a) shows the µPLE map, where resonances under quasi-resonant excitation are visible only for the lines CX1 and X. The lowest-energy CX1 resonance occurs at an excitation energy of 0.847 eV, yielding an energy difference 37 meV between excitation and emission (Fig. 2(a)). This was explained by the phonon-assisted relaxation process, including the longitudinal optical (LO) phonon, as detailed in Podemski *et al.*[28] One possible mechanism for this process is the absorption of the LO phonon (42.2 meV) and the emission of a transversal acoustic (TA) phonon (6.8 meV) in the InP matrix surrounding the QDs. The excitation energy of 0.847 eV, was selected for further studies on the dephasing time of the residual carrier spin.

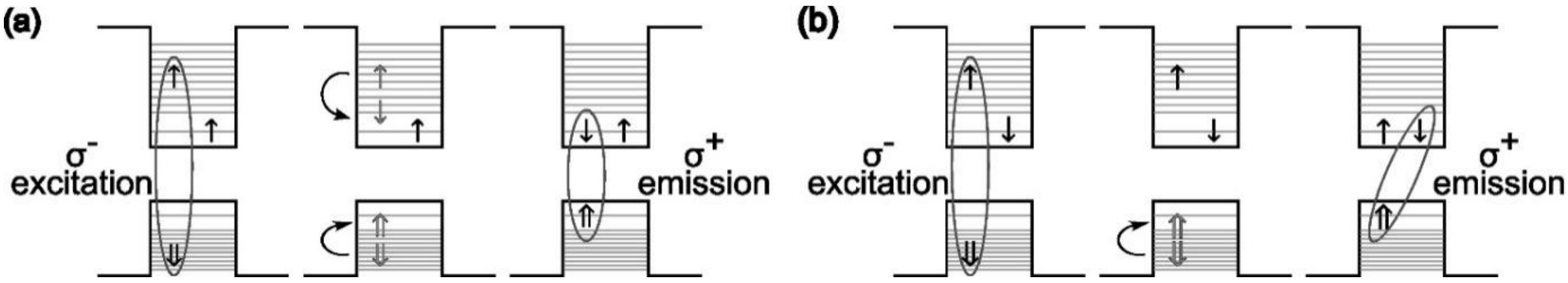

***FIG. 3***. *Mechanism of NCP creation. Excitation laser with left-circular polarization. (a) Residual and photo-created electrons with parallel spins. During relaxation, the simultaneous spin flip of electron and hole. Emission with right-circular polarization. (b) Residual and photo-created electrons with anti-parallel spins. Hole spin flip due to reduced stability, caused by weakened electron–hole exchange interaction. Emission with right-circular polarization.*

The evaluation of the spin dephasing time through the Hanle-effect measurements is based on the observation of vanishing emission polarization with increasing magnetic field strength. Here, we define the degree of circular polarization $\rho$ as:

$$\rho = \frac{I_-^- - I_+^-}{I_-^- + I_+^-} \tag{1}$$

where the $I_-^-$ ($I_+^-$) is the intensity of the emission with left (right)-circular polarization $\sigma^-$ ($\sigma^+$) under left-circularly polarized excitation laser $\sigma^-$. The $\rho > 0$ is when $I_-^- > I_+^-$, while $\rho < 0$ is when $I_-^- < I_+^-$. As shown in Fig. 2(b), the studied CX1 emission line exhibits negative circular polarization (NCP), which is characteristic of negative trions.[20,28] This confirms our predictions regarding the sign of the trions in the studied sample. The mechanisms behind the NCP creation were discussed in Masumoto *et al.*[19] There are two scenarios in which NCP can occur. With left-circular excitation polarization ($\sigma^-$), the optically generated electron has its spin ↑-oriented, and the hole has its spin ⇓-oriented. When the residual electron and the photo-generated electron have the same spin orientation (↑), due to the Pauli blocking rule, the excited electron cannot relax to the ground state (Fig. 3(a)). The spins of the photo-created electron-hole pair flip simultaneously due to the anisotropy in the exchange interaction, and relax to the ground state of the trion. Then, the electron-hole pair with the spins ↓ and ⇑, respectively, recombines with the emission of a right-circularly polarized photon ($\sigma^+$), resulting in the NCP and the residual electron being ↑-oriented. In the second scenario, where the spins of the residual and photo-created electrons are antiparallel, the spin-flip (from ⇓ to ⇑) of the hole can occur before the relaxation process (Fig. 3(b)). The total electron-pair spin is equal to zero, so the exchange interaction between hole and electron weakens, and the spin of the hole can be subject to flipping. The relaxation of the electron with the spin ↓-oriented and the hole with the spin ⇑-oriented happens with right-circularly polarized $\sigma^+$emission. The spin of the electron left in the QD is ↑-oriented, so antiparallel to the spin before excitation. Repeating the measurement many times can lead to the optical orientation of the residual electron spin. This leads to dominance of the NCP creation mechanism over the polarization-preserving process.

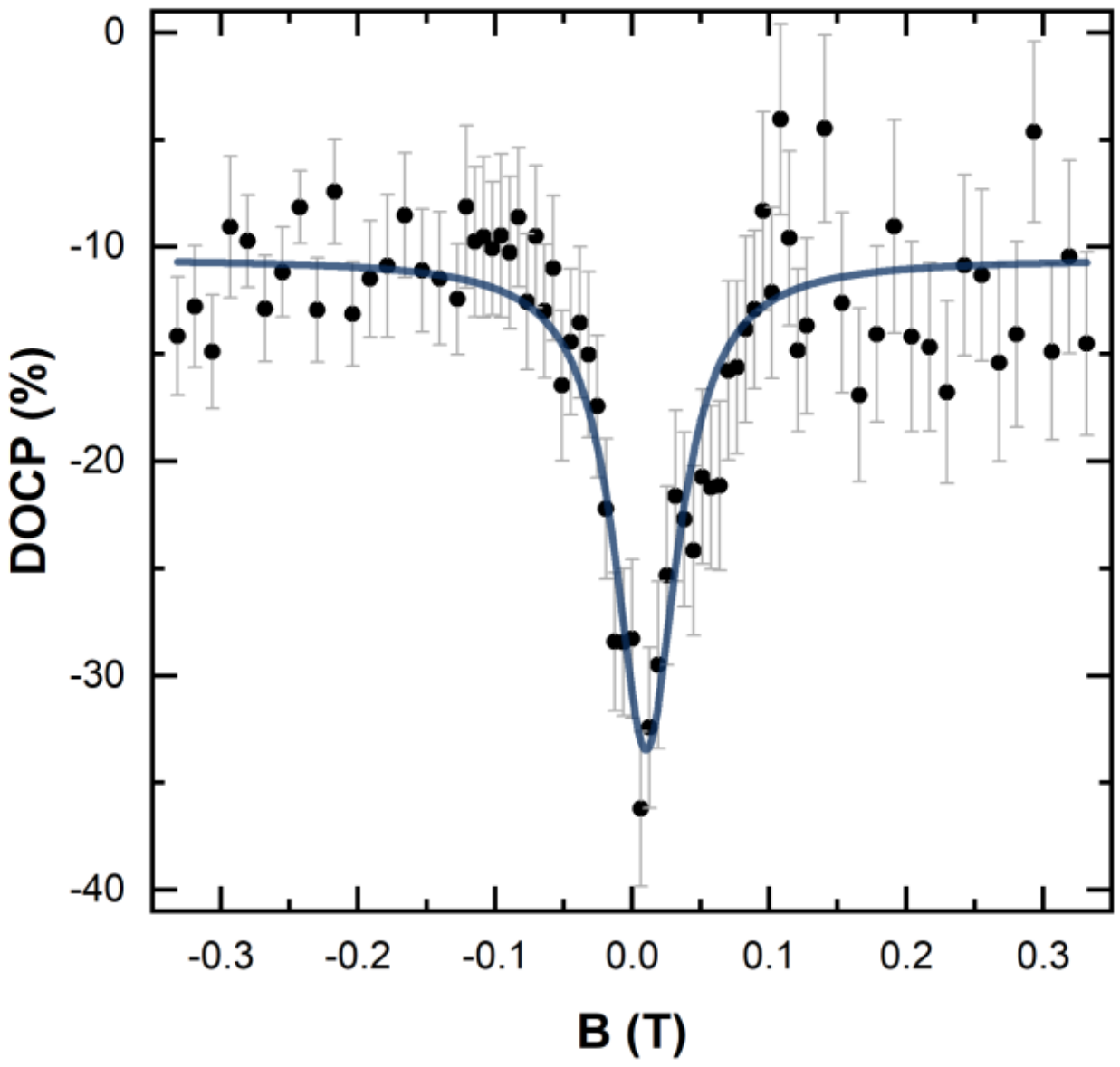

***FIG. 4.*** *Degree of circular polarization as function of external magnetic field. Obtained half width at half maximum ($B_{1/2} = 28.1 \pm 3.6\ mT$), combined with electron g-factor value known from previous study ($g_\mathrm{e} = 0.25 \pm 0.07$), corresponds to dephasing time of $T_2^* = 1.59 \pm 0.49\ ns$.*

The CX1 trion was selected for Hanle effect measurements. The obtained curve for the magnetic field in the range from -330 to 330 mT is shown in Fig. 4. The Hanle curve follows the Lorentzian shape. Here, the half-width at half maximum is given as:[20,29]

$$B_{1/2} = \frac{\hbar}{T_2^* g_\mathrm{e} \mu_\mathrm{B}} \tag{2}$$

where $T_2^*$ is the electron spin dephasing time, $g_\mathrm{e} = 0.25 \pm 0.07$ is the residual electron g-factor,[30] and $\mu_\mathrm{B}$ is the Bohr magneton. By substituting the obtained half-width of $28.1 \pm 3.6\ \mathrm{mT}$, into Equation 2, the residual electron spin dephasing time in the studied QD was estimated to be $1.59 \pm 0.49\ \mathrm{ns}$. The uncertainty of the dephasing time is determined as the compound uncertainty obtained from the standard uncertainties associated with the fitting of the emission lines for each magnetic field intensity and detection polarization, as well as the final Hanle curve fitting.

The strongest mechanism of electron spin dephasing in a QD is caused by the interaction with nuclear spins via the Overhauser field.[15-16,19] Although we study a single spin, the obtained value of the spin dephasing time is the inhomogeneous dephasing due to the many excitation/emission processes and a long measurement time.[9] However, this value is not directly comparable to measurements taken over the entire QD ensemble, where the distribution of the carriers' g-factors affects the shape of the Hanle curve. Since nuclear spins precess much slower than the electron spins, electrons experience “frozen fluctuations”, while nucleus observes the averaged long-time hyperfine field of the electrons. The effective magnetic field from the nuclear spins fluctuates, leading to the spectral diffusion and broadening of the Hanle curve. The coherence time related to the pure electron spin dephasing $T_2^*$ can be calculated using the following equation:

$$T_2^* = \hbar\sqrt{\frac{3N_{\mathrm{L}}}{16\sum_j I^j(I^j+1)\left(A^j\right)^2}} \tag{3}$$

where $N_{\mathrm{L}}$ is a number of the nuclei in the volume of the electron wavefunction (on the order of the QD size), $I^j$ and $A^j$ are the spin and hyperfine constants of the $j^{th}$ nuclei. The sum goes over the number of the nuclei in the lattice cell.[9] Studied QDs are lens-shaped with the diameter of 55 (+\-15) nm and height of 15 nm, which leads to the estimation of the number of nuclei $N = 10^6$. QDs are embedded directly in the InP matrix, causing the intermixing of the As and P atoms. We take the $I^j$ and $A^j$ for In, As, and P to be $I^{\mathrm{In}} = 9/2$, $I^{\mathrm{As}} = 3/2$, $I^{\mathrm{P}} = 1/2$, $A^{\mathrm{In}} = 47\ \mu\mathrm{eV}$, $A^{\mathrm{As}} = 44\ \mu\mathrm{eV}$, and $A^{\mathrm{P}} = 36\ \mu\mathrm{eV}$. [31-32] Based on these values, we estimated the $T_s \sim 1$ ns, which is consistent with our result and indicates that the measured process of the spin dephasing is driven by the Overhauser field through the hyperfine interaction. From Equation 3, we calculated that when changing the intermixing of the P with As atoms from 0 to 100%, the dephasing time increases only slightly (by around 0.03 ns). As spins and the hyperfine constants of P and As atoms nuclei are comparable, the intermixing does not affect strongly the resulting electron spin dephasing time. Additionally, P atoms are located close to QD edge, where the electron wave function is decreasing, so the overlap integral with nuclear spins is negligible. On the other hand, the spin dephasing time depends more on the volume of the QD, as it increases proportionally to $\sqrt{N_{\mathrm{L}}}$.

The electron spin dephasing time observed in our InAs(P)/InP QDs is comparable to previous reports for III–V semiconductor QDs emitting at wavelengths around or below 1 µm. For self-assembled InAs/GaAs QDs grown by MBE, both experimental and theoretical studies report scaled dephasing times $gT_2^* \sim 100\ \mathrm{ps}$[9,15] (in our measurements $gT_2^* = 400\ \mathrm{ps}$). For InP/$In_{0.5}Ga_{0.5}P$/GaAs QD ensembles grown by MBE, Masumoto *et al*. report $gT_2^* = 2.5\ ns$, corresponding to a dephasing time $T_2^* = 1.7\ \mathrm{ns}$ for $g = 1.5$.[19] GaAs/AlGaAs/GaAs QDs grown via nanohole droplet etching are expected to have longer coherence times due to the low strain and absence of In. However, experimentally determined dephasing times in those QDs have only reached 2-4 ns.[14,33] For QDs emitting at the telecom wavelengths, the obtained results are also comparable. Study on InAs/$In_{0.53}Al_{0.24}Ga_{0.23}As$/InP QDs ensemble revealed dephasing time of electrons spins $T_2^* = 0.6\ \mathrm{ns}$,[34] while for single QDs the electron spin dephasing times tend to be on the level of 1 ns, for both InGaAs/GaAs,[22] and InAs/InP QDs.[23]

A key material-dependent factor in spin dephasing is the nuclear spin composition. Indium, with a nuclear spin $I^{\mathrm{In}} = 9/2$, dominates the hyperfine interaction in In-containing systems, leading to faster dephasing. In contrast, in GaAs QDs with $I^{\mathrm{Ga}} = 3/2$, a weaker interaction is expected. Although P atoms are incorporated in the studied QDs, the intermixing of the As and P atoms occurs mostly at the border of the QDs, where the overlap of the electron and atomic nuclei wave functions is relatively weak. Therefore, the calculated electron spin dephasing time shows that the QD geometry plays a crucial role. In larger QDs the electron wavefunction is more spatially extended, which reduces its overlap with the atomic nuclei. This weaker interaction suppresses hyperfine-mediated decoherence. This effect has previously been observed in InGaAs/GaAs QDs.[35] Despite the presence of indium in our system, which tends to shorten spin coherence due to its high nuclear spin and strong hyperfine coupling, the relatively large QD size positively contributes to the observed dephasing time.

We studied the spin dynamics of the single electron confined in the InAs(P)/InP QD using the Hanle effect measurement. The DOCP of the emission line originating from the negative trion recombination exhibited −36% in zero magnetic field and followed the Lorentzian shape with the magnetic flux density. Based on the Lorentzian half-width, we estimated the dephasing time of the single spin of the resident electron to be $1.59 \pm 0.49$ ns. In external magnetic fields weaker than the Overhauser field, the electron spin is affected by the so-called "frozen fluctuation" of the nuclear spin, so the result suggests that the dephasing is caused by the hyperfine interaction. Despite the large In spin, the relatively large size of telecom QDs elongates the spin dephasing time. This makes the studied QDs attractive candidates for spin qubit applications compared to those in other material systems.

## Acknowledgments

This work was co-financed by the Ministry of Education and Science, Republic of Poland within the „Perły Nauki" project, grant no. PN/01/0117/2022. This research was funded by the Foundation for Polish Science co-financed by the EU under the ERDF by project entitled „Quantum dot-based indistinguishable and entangled photon sources at telecom wavelengths" carried out within the HOMING program. This work was also financially supported by the Bundesministerium für Forschung, Technologie und Raumfahrt-BMFTR in the frame of the project QR.N (16KIS2204) and DFG-Heisenberg grant (BE 5778/4-1). We acknowledge Andrei Kors for his assistance in the MBE growth process, Kerstin Fuchs and Dirk Albert for technical support.

## AUTHOR DECLARATIONS SECTION

### Conflict of Interest

The authors have no conflicts to disclose.

### Author Contributions

**M.W.:** Investigation (lead); Formal Analysis (lead); Visualization (lead); Writing – original draft (lead); Writing – review and editing (equal); Funding Acquisition (equal). **H.J.:** Investigation (supporting); Formal Analysis (supporting); Visualization (supporting). **A.M.:** Methodology (supporting); Writing – original draft (supporting); Writing – review and editing (equal). **J.P.R:** Resources (equal); Writing – review and editing (equal); Funding Acquisition (equal); Supervision (lead). **M.B.:** Resources (equal);

Writing – review and editing (equal); Funding Acquisition (equal). **W.R.-R.:** Conceptualization (lead); Methodology (lead); Writing – original draft (supporting); Writing – review and editing (equal).

## DATA AVAILABILITY

The data that support the findings of this study are available from the corresponding author upon reasonable request.